\documentclass[12pt]{article}

\usepackage{scicite}
\usepackage{setspace}
\usepackage{caption}
\DeclareCaptionLabelFormat{bold}{\textbf{#1 #2}}
\usepackage{times}
\usepackage{subfiles}
\usepackage{amsmath}
\usepackage{graphicx} 
\usepackage[colorlinks = true,
            linkcolor = blue,
            urlcolor  = blue,
            citecolor = blue,
            anchorcolor = blue]{hyperref}

\newenvironment{sciabstract}{%
\begin{quote} \bf}
{\end{quote}}

\newcounter{lastnote}

\title{Environmental Burden of United States Data Centers in the Artificial Intelligence Era}

\author
{Gianluca Guidi,$^{1,2}$ Francesca Dominici,$^{1\dagger}$ Jonathan Gilmour$^{1}$, Kevin Butler,$^{3}$\\ Eric Bell,$^{4}$ Scott Delaney$^{5}$, Falco J. Bargagli-Stoffi$^{1,6}$\\
\\
\normalsize{$^{1}$Department of Biostatistics, Harvard T.H. Chan School of Public Health,}\\ \normalsize{Boston, Massachusetts, USA
}\\
\normalsize{$^{2}$Department of Computer Science, University of Pisa, Pisa, Italy}\\
\normalsize{$^{3}$Environmental Systems Research Institute, Redlands, California, USA}\\
\normalsize{$^{4}$Baxtel, Denver, Colorado, USA}\\ 
\normalsize{$^{5}$Department of Environmental Health, Harvard T.H. Chan School of Public Health,}\\ 
\normalsize{Boston, Massachusetts, USA}\\
\normalsize{$^{6}$Department of Biostatistics, UCLA Fielding School of Public Health}\\ \normalsize{Los Angeles, California, USA}\\
\\
\normalsize{$^\dagger$To whom correspondence should be addressed; E-mail:  fdominic@hsph.harvard.edu.}
}

\date{}

\begin{document} 


\baselineskip24pt


\maketitle 

\vspace{-1cm}

\singlespacing
\begin{sciabstract}
The rapid proliferation of data centers in the US---driven partly by the adoption of artificial intelligence---has set off alarm bells about the industry’s environmental impact. We compiled detailed information on 2,132 US data centers operating between September 2023 and August 2024 and determined their electricity consumption, electricity sources, and attributable CO$_{2}$e emissions. Our findings reveal that data centers accounted for more than 4\% of total US electricity consumption---with 56\% derived from fossil fuels---generating more than 105 million tons of CO$_{2}$e (2.18\% of US emissions in 2023). Data centers' carbon intensity---the amount of CO$_{2}$e emitted per unit of electricity consumed---exceeded the US average by 48\%. Our data pipeline and visualization tools can be used to assess current and future environmental impacts of data centers.
\end{sciabstract}

{\noindent \textbf{One sentence summary:} US data centers produced 105 million tons CO$_{2}$e in the past year with a carbon intensity 48\% higher than the national average.}
\thispagestyle{empty}
\clearpage

\newpage

\pagenumbering{arabic}
\setcounter{page}{1}

\section*{Introduction}

\doublespacing

Data centers---warehouses containing thousands or millions of computing cores that serve as the backbone of modern information technology infrastructure---provide the necessary resources and environment to process, store, and distribute vast amounts of data, playing a critical role in supporting a wide range of activities, including cloud computing, online services, e-commerce, social media, scientific research, and more \cite{kamiya2024energy}. 

Data centers are energy-intensive facilities, with computational power and cooling as the most energy-hungry processes \cite{sun2021prototype}. Data center servers, depending on the task, require substantial energy to perform their computations, and this computing process can generate significant amounts of heat. Extensive cooling systems are required to avoid computer hardware overheating and maximize performance, stability, and life expectancy, especially in high-performance systems \cite{dayarathna2015data}. 

The increasing demand for digital services has made data centers essential and widespread in modern society \cite{crawford2024generative,erdenesanaa2023ai,patterson2021carbon}.In 2024, there were more than 7,945 data centers worldwide \cite{DataCenterMap24}. Of these, 2,990 are located in the US (37\%) \cite{DataCenterMap24}. Globally, according to the International Energy Agency (IEA), energy consumption of data centers in 2022 was estimated to be between 240 and 340 terawatt hours (TWh), which represents approximately 1-1.3\% of global electricity consumption \cite{iea_datacentres}. Data center energy consumption is projected to double by 2026, reaching 480-680 TWh worldwide \cite{iea_datacentres}. This exceeds the energy consumption of Canada, the country with the sixth highest energy consumption in the world \cite{enerdata_electricityconsumption}. 

In the US alone, data center electricity consumption is projected to increase from approximately 200 TWh in 2022 to nearly 260 TWh in 2026 \cite{thenewstack_datacenters_energy}. This increase means that data centers will account for 4\% to 6\% of the total US electricity consumption by 2026 \cite{thenewstack_datacenters_energy}. More recently, the exponential rise of artificial intelligence (AI) and the widespread adoption of generative AI models have profoundly impacted the need for data centers, significantly increasing demand in terms of number and size. This has exacerbated the strain on electricity resources and triggered an unprecedented increase in energy demand and consumption \cite{stokel-walker2024generative,strubell2019energy}.

Electricity generation has a significant environmental impact, with various forms of generation affecting air quality, water resources, and ecosystems \cite{epa2024electricity}. This is particularly true for carbon-intensive energy sources, those that produce a large amount of carbon dioxide equivalent (CO$_{2}$e) emissions per unit of electricity produced, such as coal and natural gas. CO$_{2}$e encompasses all greenhouse gases that contribute to climate change, providing a standardized measure to assess environmental impacts.

A recent study by Siddik et al. estimated that data centers in the US emitted 31.5 million tons of greenhouse gases in 2018 \cite{siddik2021environmental}. Unfortunately, estimates of the carbon footprint of data centers for 2018 are already outdated. Additionally, Siddick et al. provide estimates of the nine large geographic regions defined by NOAA, a level of granularity that does not align with possible regulatory responses.

In this paper, we introduce a data pipeline to estimate the environmental impact of US data centers up to August 2024 and for 52 balance authority regions. A data pipeline is an automated process that collects, processes, and analyzes information about energy consumption and emissions to estimate and monitor the environmental impact of a data center's activities. Because it is automated, it can be routinely updated and leveraged to assess the effectiveness of interventions to reduce energy consumption and/or rely on alternative sources of energy. A balancing authority area in the US is a specific geographic region within which a single entity---known as a balancing authority---is responsible for maintaining the balance between electricity generation and consumption, as well as managing power flows with neighboring areas.

More specifically, we gathered detailed data on 2,132 US data centers (78\% of all US data centers, \cite{DataCenterMap24}) for the period from September 2023 to August 2024. Cryptocurrency mining data centers were excluded, as they have been extensively studied in Guidi et al. \cite{guidi2024environmental}. For each of the data centers included in our analysis, we: 1) identified and validated their energy demand; 2) pinpointed the power plants that supply the electricity to each of the data centers; 3) identified each individual power plant's share of electricity supplied and the fuel used to produce the electricity; and 4) estimated the CO$_{2}$e emissions. 

We also estimated and compared CO$_{2}$e emissions and \textit{carbon intensities}---defined as the amount of CO$_{2}$e produced per unit of energy consumed---for data centers versus other relevant sectors of the US economy and those of other countries. 

Finally, we developed a public-facing web platform to track CO$_{2}$e emissions, energy load estimates, and total energy demand of data centers at the balancing authority region and at the state level.  The data science pipeline introduced in this paper provides novel visualization tools and a platform to track and mitigate the carbon footprint and assess the current and future environmental impact of data centers.

\section*{Results}

\subsection*{Characteristics of Data Centers}

We identified 2,132 data centers located throughout the contiguous US that were operating from September 2023 to August 2024 (Fig. \ref{fig:location}). Virginia had more data centers (301) than any other state, followed by California (248) and Texas (221).

We gathered detailed information for each data center, including full address, exact location of the building, square footage, and power capacity specifications. Given the lack of open source datasets for data centers, we built our data pipeline by integrating publicly available data, including data obtained by Web scraping, with proprietary data from the data center service provider (\textit{Baxtel.com}). We validated the data set through satellite imagery using \textit{Open Street Map}. Additional details on data collection and validation are provided in the Supplementary Materials.

\begin{figure}
\begin{center}
\includegraphics[width=0.75\textwidth]{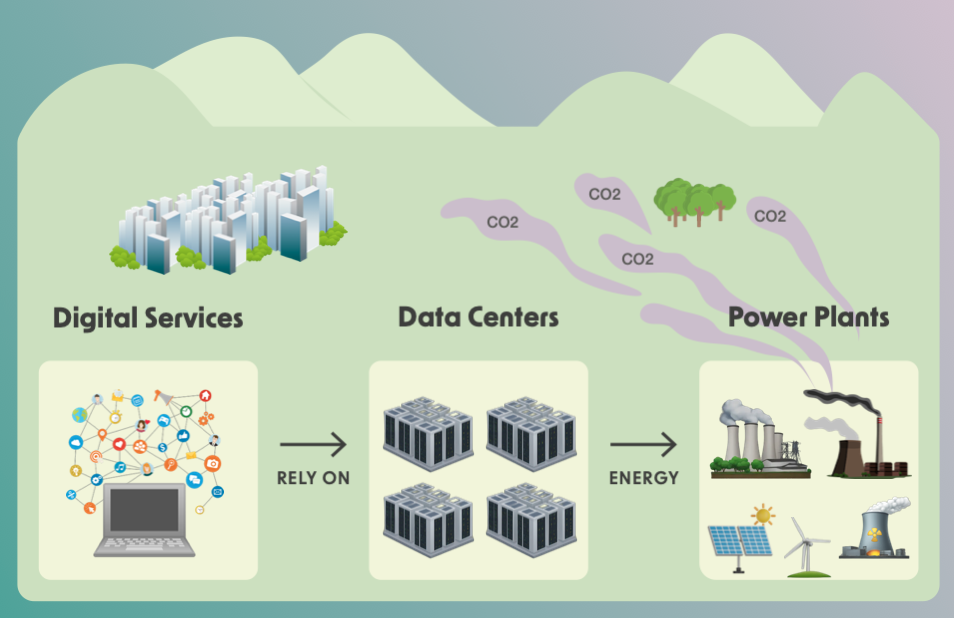}
\end{center}
\caption{\textbf{The chain of digital services, data centers, and CO$_{2}$ emissions.} This infographic illustrates the interconnected relationship between digital services, data centers, energy consumption, and the consequent emissions from the power plants that supply data centers. It highlights how data centers power digital services by relying on energy from power plants, leading to CO$_2$e emissions that are subsequently released into the environment.}
\label{fig:diagram}
\end{figure}

\subsection*{Electricity Consumption of Data Centers}

We obtained power capacity specifications for 1,795 data centers (84\% of the total sample). Power capacity refers to the maximum amount of electricity that can be drawn by each data center, expressed in megawatts (MW). For the 337 data centers with missing power capacity data, we deployed a gradient-boosted regression tree model to estimate their power capacities. We trained the model using features from the 1,795 data centers with complete data and then estimated the power capacity of the remaining data centers. 

The input of the model included data center features such as square footage, location and regional climate, and hyperscale capability, that is, whether the data center was designed for extreme scalability and optimized for large-scale workloads with enhanced network infrastructure \cite{ibm_hyperscale_datacenter}. The model for imputation of missing data had an R-squared of 0.77 (see Supplementary Materials for details).

We found an estimated aggregate electricity consumption of 192.64 Twh for the 2,132 data centers during the study period, assuming a constant uptime, or capacity utilization rate, of 0.75 throughout the year, which represents the percentage of time a data center is expected to operate at its maximum capacity (see Supplementary Materials for details). The US IEA estimated that in 2022, the total electricity consumption of data centers in the US was 200 TWh, signaling that our sample, which captured data from most of the data centers in the US, was slightly more conservative relative to the estimate of the IEA of 2022 \cite{iea_datacenters_networks}.

The aggregated energy consumption of all data centers included in our analysis (192.64 TWh) represented approximately 4.59\% \cite{eia_energyconsumption} of the total energy consumption in the US in 2022 and was higher than that of most US states, ranking just after Texas, California, and Florida (at 475, 251 and 248 TWh, respectively) \cite{EIA_fuel_use}. This figure is more than double the 1.80\% estimate reported for 2018 \cite{siddik2021environmental}.

Furthermore, the aggregate energy consumption of data centers in our study was comparable to that of developed countries, such as Spain (220 TWh) and Italy (298 TWh), and was considerably higher than some Scandinavian countries (Norway and Sweden, both around 124 TWh) and Argentina (134 TWh) \cite{enerdata_domestic_consumption} (see Supplementary Materials for more details). Aggregate energy consumption was also slightly lower than an estimate for 2018 for the total global energy use of data centers of 205 TWh \cite{masanet2020recalibrating}.

The data centers in our sample ranged in power capacity from 0.04 to 325 MW, with a mean of 13.75 MW and a median of 4.5 MW. Data centers in Virginia had the highest aggregate power capacity (and consequently electricity consumption) among all US states, with 52.21 TWh, which was more than 27\% of the total electricity consumed by all data centers in our dataset. Texas was second with 18.86 TWh (13\% of the total), followed by Oregon, with 15.25 TWh (8\%) and California with 11.54 TWh (6\%). These four states accounted for more than 50\% of the total energy consumption of the data centers in our study.

\subsection*{Power Plant Production to Meet the Energy Demand of the Data Centers}

Energy production is regulated within a balancing authority region and not at the data center level \cite{epa_egrid_faq}. Each region is a well-defined geographical area within the electric grid system that is managed by a single balancing authority \cite{epa_egrid_faq}. 

To identify the power plants that supply electricity to each data center, we first assigned each data center to its balancing authority based on the data center's geographical location and then linked each data center to the power plants in that balancing authority region. Second, we implemented a ``generation-weighted average'' approach to apportion the electricity consumed by each data center to each supplying power plant. We assumed that each power plant within a balancing authority region supplied to a given data center an amount of energy proportional to the energy it generated in that year. 

Using a generation-weighted average is an \textit{attributional} approach to CO$_{2}$e accounting \cite{brander_most_2022}. Attributional methods measure a static inventory of emissions allocated proportionally to users and are useful for setting carbon budgets and setting CO$_{2}$e reduction targets. Given the absence of comprehensive industry-wide reporting on emissions from the data center sector, we employed attributional methods to estimate the aggregate CO$_{2}$e emissions. 

We identified 3,318 power plants across the US that supplied electricity to the 2,132 data centers in our dataset. Fossil fuel power plants produced more than 56\% of the consumed by data centers.

\subsection*{Carbon Emissions Attributable to Data Centers}

After identifying the power plants that generate electricity for each data center, we estimated the corresponding aggregate CO$_{2}$e emissions attributable to data center energy consumption.

\begin{figure}
\begin{center}
\includegraphics[width=1\textwidth]{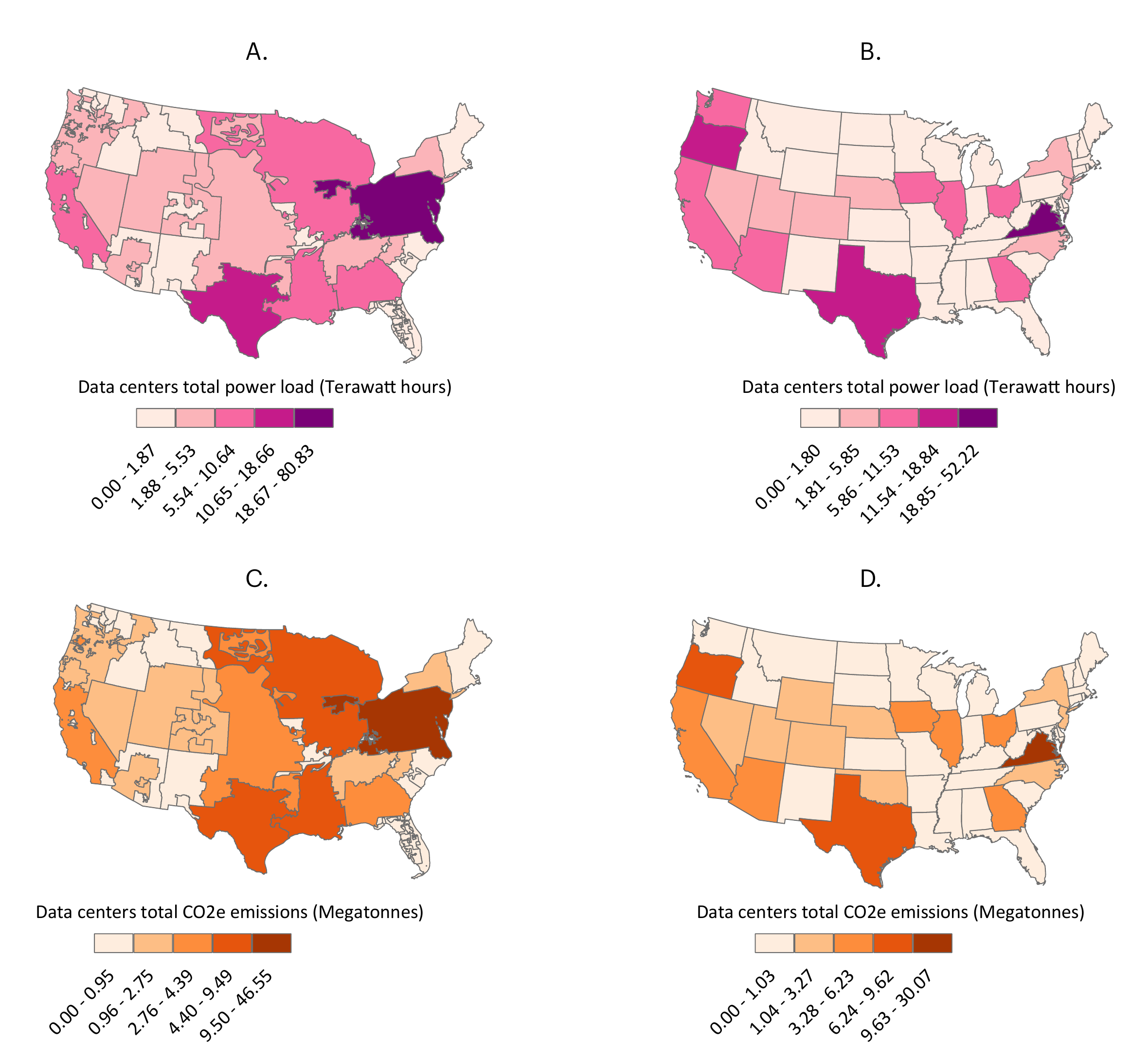}
\end{center}
\caption{\textbf{Data center energy consumption and CO$_{2}$ emissions.} (Left column, A and C) The balancing authority region in which a data center is located determines the mix of power plants that supply its electricity and thus its attributable emissions. (Right column, B and D) Maps at the state level show energy consumption and emissions for which the data centers within the state are responsible for.}
\label{fig:powerload_emissions}
\end{figure}

The total CO$_{2}$e emissions of the data centers in our data set was 105.59 million metric tons (MT), equal to 2.18\% of US carbon emissions from energy consumption \cite{EIA_carbon_emissions}, and to 1.66\% of US total CO$_2$e emissions in 2022 \cite{EPA2023}. This is more than three times the proportion reported for 2018 in \cite{siddik2021environmental}, of 0.5\%. We found that Virginia, Texas, and Oregon had the highest CO$_{2}$e emissions attributable to data centers: 30.08, 9.63, and 8.92 MT of emissions, respectively (see Table \ref{tab:table1}). Illinois was fourth with 6.22 MT and California was seventh with 4.37 MT of CO$_{2}$e emissions. Fig. \ref{fig:powerload_emissions} shows the distribution of energy loads and emissions across the balancing authority regions and at the state level.

\begin{table}[h!]
\begin{center}
\includegraphics[width=0.95\textwidth]{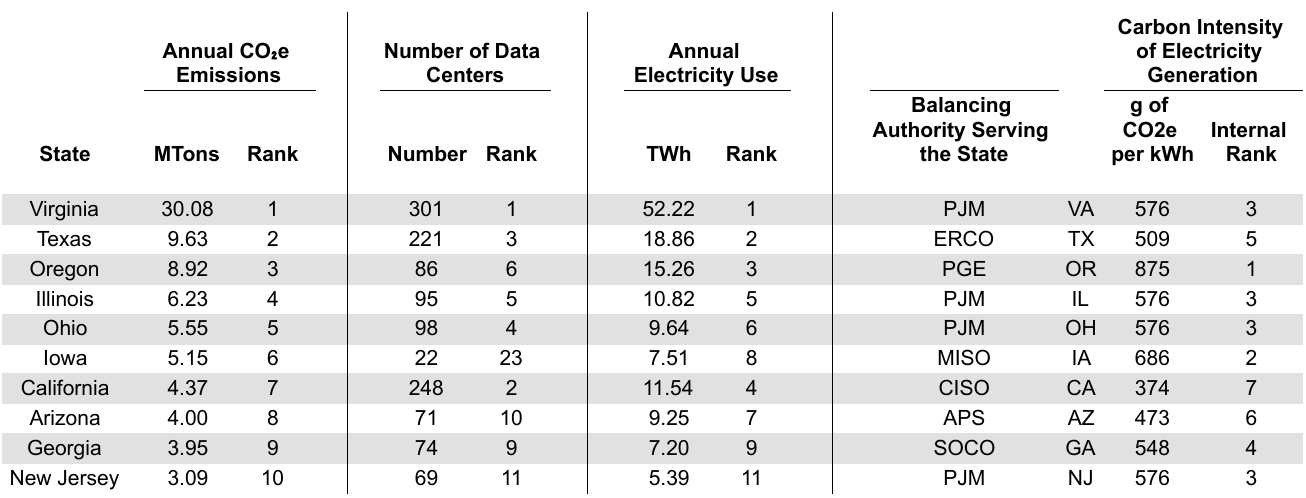}
\end{center}
\caption{\textbf{Statistics for the top ten states by CO$_{2}$e emissions attributable to data center energy consumption}. Annual electricity use assumes a constant 0.75 uptime across all data centers. The emissions and annual electricity displayed correspond to those attributable to data centers within the states. However, due to the role of balancing authorities, emissions do not necessarily coincide with the locations of the data centers. The balancing authorities shown are the ones primarily supplying energy to the states. See Fig. S.1.6 for balancing authorities in the US. The \textit{Internal Rank} column depicts the rank between the balancing authorities. Note that this ranking is computed only among the balancing authorities reported in the table.}
\label{tab:table1}
\end{table}

\subsection*{Carbon Intensity and Energy Fuel Mix of Data Centers}

\begin{figure}
\begin{center}
\includegraphics[width=1\textwidth]{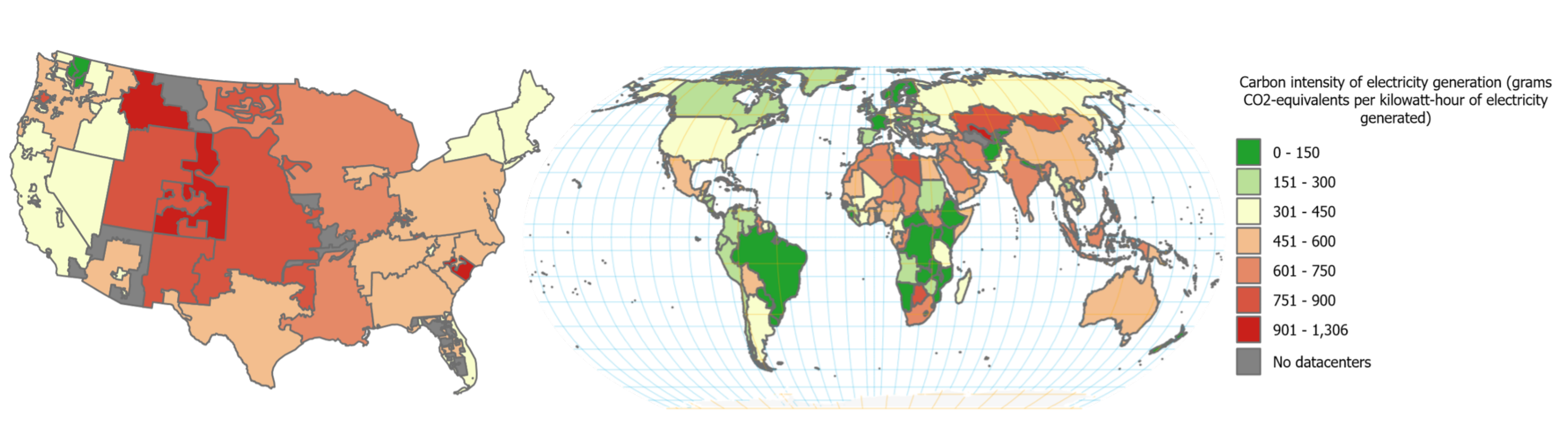}
\end{center}
\caption{\textbf{Carbon intensities of energy production for US data centers by balancing authorities and in world countries.} The left panel shows data centers' carbon intensity for energy production at the balancing authority level, in grams of CO$_{2}$e per kWh. The right panel shows the carbon intensities of total energy production by world countries in grams of CO$_{2}$e per kWh.} 
\label{fig:carbon_intensities}
\end{figure}

Next, we evaluated the \textit{cleanliness} of the energy supplied by the power plants to data centers, measured through their carbon intensity. Fig. \ref{fig:carbon_intensities} shows the carbon intensities of the US data centers at the level of the balancing authority. To put these numbers in context, we also determined the carbon intensities of the entire power generation sector (not only of data centers) of other countries.

The average carbon intensity of the US data centers in our study (weighted by the energy they consumed) was 548 grams of CO$_{2}$e per kilowatt hour (kWh), approximately 48\% higher than the US national average of 369 gCO$_{2}$e / kWh \cite{ourworldindata_carbon_intensity}. This strikingly high percentage suggests that data centers in the US are located in geographical areas where energy sources are more carbon-intensive than the national average. 

Fig. \ref{fig:carbon_intensities}, left panel, shows carbon intensities at the level of the balancing authority. Central balancing authority regions, such as those in Colorado, Kansas, Missouri, and Wyoming, had the highest carbon intensities, due to their heavy dependence on coal-fired power plants \cite{henneman2023mortality}, with peak values around 1000 gCO$_{2}$e/kWh. 

Virginia's balancing authority (PJM), despite having the largest number of data centers, reported a lower carbon intensity (576 gCO$_{2}$e/kWh). However, the carbon intensity of this region was significantly above the national average (369 gCO$_{2}$e / kWh), which increased the weighted average carbon intensity of the data centers. 

The balancing authority serving Texas (ERCO) had the second highest aggregate data center electricity demand and a carbon intensity of 509 gCO$_{2}$e/kWh. The third highest carbon intensity was observed for MISO, which serves multiple states, including Illinois and Iowa (fourth and sixth in CO$_{2}$e emissions attributable to data centers), with a carbon intensity of 686 gCO$_{2}$e/kWh. 

\begin{figure}
\begin{center}
\includegraphics[width=1\textwidth]{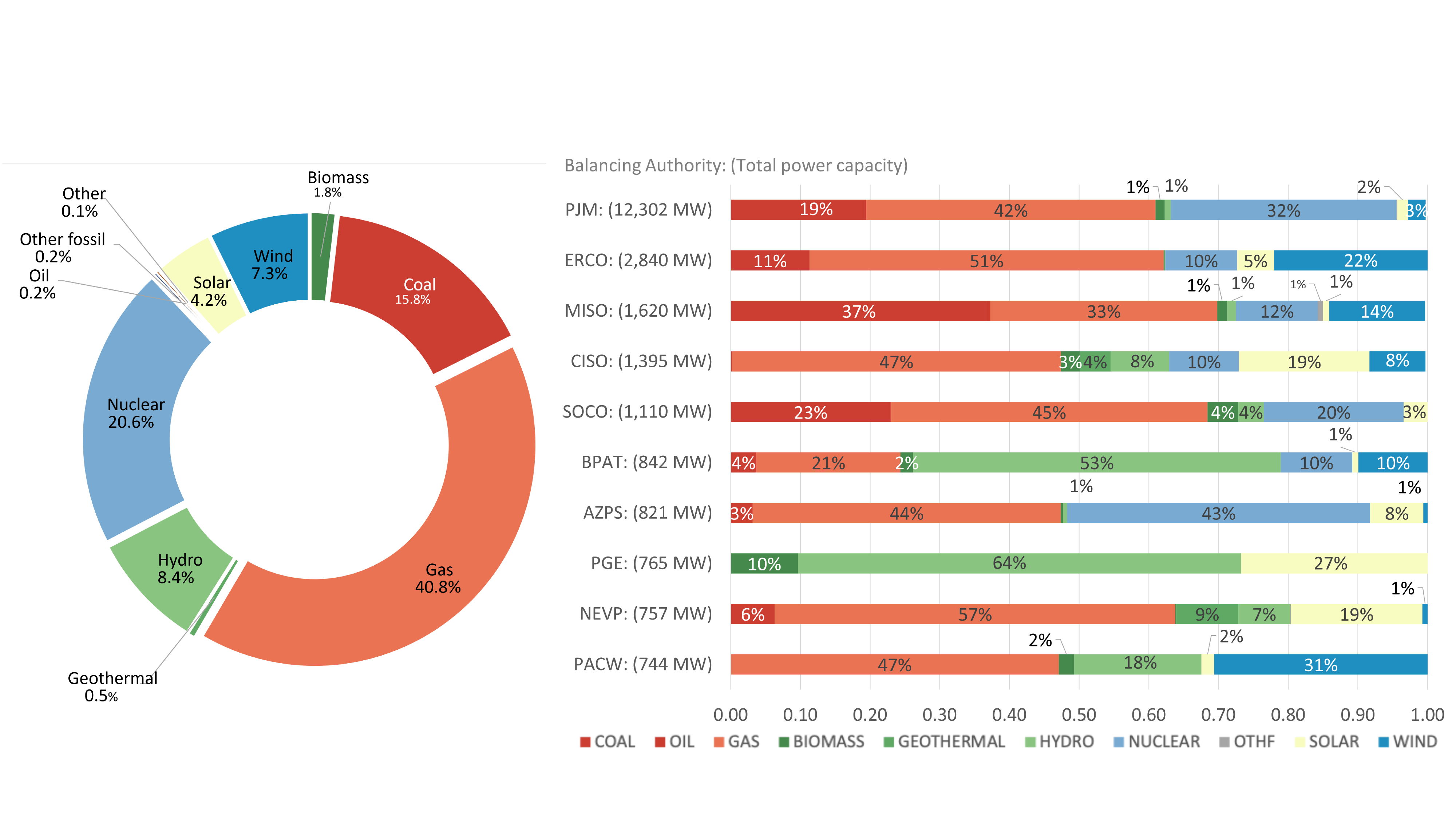}
\end{center}
\caption{\textbf{Fuel mix of power plants producing energy for US data centers.} The left panel depicts the distribution of fuel types used by the power plants providing energy for US data centers in our study. The right panel shows the first 10 balancing authorities ranked by aggregated power capacity of data centers in the balancing authority region (shown on the vertical axis), and the amount of energy produced per fuel type.}
\label{fig:fuel_mix}
\end{figure}

Finally, CISO, California's main balancing authority, had the fourth highest aggregate data center electricity capacity, but a much lower carbon intensity of 373 gCO$_{2}$e/kWh. California has almost completely phased out coal power plants and relies on a split between natural gas and sustainable energy sources \cite{california_energy_commission} (see Fig. \ref{fig:fuel_mix}).

Comparing the weighted average carbon intensity of US data centers (weighted by their energy consumption) with the total carbon intensities of various countries (across all sectors and not only for the data centers), we found that it was similar to that of China (582 gCO$_{2}$e/kWh), Australia (549 gCO$_{2}$e/kWh), Mexico (507 gCO$_{2}$e/kWh), and Bolivia (532 gCO$_{2}$e/kWh). 

Moreover, US data center average carbon intensity was significantly higher than the carbon intensity of several European countries, including France (58 gCO$_{2}$e/kWh), Spain (174 gCO$_{2}$e/kWh), Italy (331 gCO$_{2}$e/kWh), the United Kingdom (238 gCO$_{2}$e/kWh), and Germany (381 gCO$_{2}$e), as well as countries such as Russia (441 gCO$_{2}$e/kWh), Argentina (354 gCO$_{2}$e/kWh), and Brazil (98 gCO$_{2}$e/kWh) \cite{ourworldindata_carbon_intensity} (see Fig. \ref{fig:carbon_intensities}, right panel).

Carbon intensities are closely related to the fuel mix used in energy production, which varies significantly depending on the country, region, and types of power plants. The primary components of the fuel mix include fossil fuels (coal, natural gas, and oil), renewable energy (solar, wind, hydro, geothermal, and biomass), and nuclear power. 

Fig. \ref{fig:fuel_mix} illustrates the fuel dependence of all US data centers in our dataset and the main balancing authorities by power capacity within these regions. As the left panel shows, fossil fuels dominated, accounting for more than 56\% of all the electricity generated for US data centers. Nuclear power provided approximately 21\% of the energy consumed by data centers, while renewable energy provided about 22\%. balancing authority with the most data centers and highest electricity use—serving Virginia, 12 other states, and the District of Columbia, more than 60\% of the electricity came from fossil fuel power plants, with coal-fired plants accounting for 19\% of the total. Nuclear represented 32\% of the energy mix, and renewables a small 7\%.
 
\subsection*{Publicly Available Web Application}

We developed a data visualization tool that allows users to track the carbon footprint of US data centers. The platform can be found at \texttt{\url{https://tinyurl.com/4k8fbhka}}. The platform offers different geographical layers---balancing authority, state, and county levels---at which the number of data centers and their energy consumption, attributable carbon emissions, carbon intensities, and fuel mixes can be explored.

\section*{Discussion and Policy Implications}
 
The environmental impact of the rapid proliferation of data centers is obfuscated by the relative unavailability of industry emissions data. To close this critical data gap and to enable effective policymaking, we have developed a reproducible, data-driven approach that reveals data centers' energy consumption and attributable emissions. 

Our analysis reveals that data centers were responsible for 105 million tons of carbon dioxide equivalent emissions attributable in the past year, a three-fold increase in carbon emissions compared to the 2018 estimates by Siddik et al. \cite{siddik2021environmental}. 

We estimate these emissions at the balancing authority level (52 regions responsible for energy generation) and at the state level (where regulatory actions are implemented). This granularity aligns energy consumption and emissions with the geographical areas responsible for energy production and policy, offering an actionable foundation for regulatory decisions.

We also provide a breakdown of the fuel mix that powers the data centers by balancing authority. Alarmingly, we found that approximately 56\% of the electricity consumed by the data centers came from fossil fuel-burning power plants, and 16\% was from coal power plants. This high dependence on coal is a significant contributor to the carbon footprint of the data center industry. 

In the midst of the AI race, various companies have fallen off the track with their sustainability pledges. Google’s 2023 greenhouse gas emissions increased 13\% year-over-year \cite{google_sustainability2024}. Microsoft's emissions have increased 29\% since their 2020 baseline \cite{microsoft_sustainability2024}. Both cited data centers as the main driver of the increase in these emissions.

Our analysis and these industry trends underscore the importance of recognizing data centers as a distinct sector of the economy. Policy has lagged behind the explosive growth of data centers, though states and federal regulators are increasingly turning their eye to the industry \cite{datacenter_knowledge_tougher_reporting, uptimeinstitute_mandates_crypto}. 

We hypothesize that the carbon intensity of electricity is often overlooked when a new data center is built. In fact, we found that about 95\% of data centers are located in areas with a higher carbon intensity of electricity than the national average, despite the fact that some of the biggest data center operators are also the largest private purchasers of renewable energy \cite{bnef_clean_energy}. 

Our analyses underscore the importance of relying on data-driven decision-making to mitigate the environmental impact of data centers while acknowledging their crucial role as the backbone of modern computing technology. The urgency of this matter is further amplified by the anticipated surge in data center proliferation driven by the rapid advancement and widespread adoption of AI technologies. The AI race significantly increases computational demands and, consequently, energy consumption of data centers.

\subsection*{Limitations and Future Research}

Despite incorporating a large dataset that covers most data centers in the US and applying conservative assumptions, our study has limitations. First, the uptime, or capacity utilization rate, is assumed to be 0.75 for all data centers. The uptime impacts data center energy consumption computations and therefore emissions levels. A more precise model would factor in a specific uptime for each data center, accounting for the varying types (e.g. hyperscale, colocation, etc.). However, this data is closely guarded by the data center industry and is not readily available.

Furthermore, our analysis of power plants by balancing authority was limited to the 3,318 facilities for which power generation and emissions data were available from the EPA's eGRID database, out of a total of 11,973 reported power plants. This limitation arises from EPA reporting thresholds, which require data submission only from power plants exceeding certain sizes or emissions levels \cite{epa_egrid_faq}. Smaller or lower-emission plants may not be required to report. While this may slightly affect the completeness of our balancing authority assignments— and thus, carbon intensity and fuel mix assessments- we anticipate minimal impact on our results, as the excluded plants are generally quite small.

Lastly, when estimating power capacities for facilities with missing data, we primarily relied on the square footage of the data centers, as indicated in the literature, as well as their geographic location and climate zone. Factors such as industry economies of scale and regional temperature profiles were considered as potential determinants of power capacities. Although these methods provided reasonable estimates, more accurate predictions could be achieved by incorporating additional facility-specific information.

Future research should focus on developing methodologies to identify new data center facilities as they come online. This would involve both the refinement of data collection techniques and the establishment of collaborative networks with public and private stakeholders to ensure timely and accurate reporting of new facilities. 

In this paper, we adopted an \textit{attributional approach} (i.e., a generation-weighted average emission model), to assign each data center the power plant that supplied the energy demanded. Other approaches, called \textit{consequential},  have been used in the literature on emission accounting \cite{xing2023carbonrespondercoordinatingdemand,jagannadharao2023timeshifting,dodge2022measuringcarbonintensityai}, and serve a different purpose, that is, identify power plants that respond to the \textit{increase} in energy demand from the establishment of new energy-hungry facilities (e.g., a new data center)  \cite{Gagnon_and_Cole2022}. 

These approaches measure the change in total emissions from a change in energy demand \cite{brander_most_2022} (i.e., the marginal emissions rate).  The responding, or marginal, power plants in these situations generally do not have the same emissions rates as the average of all existing power plants (consequential approach).  

When making decisions on when or where to operate a new data center, it will be important to use consequential methods that model the change in total emissions as a consequence of an intervention. A preliminary consequential analysis is presented in the Supplementary Material. However, a more comprehensive approach is left for future research.

\bibliographystyle{Science}
\bibliography{scibib}

\begin{thebibliography}{10}

\bibitem{kamiya2024energy}
G.~Kamiya, P.~Bertoldi.
\newblock {Energy consumption in data centres and broadband communication networks in the EU}.
\newblock {\it European Commission, Joint Research Centre\/}  (2024).

\bibitem{sun2021prototype}
K.~Sun, N.~Luo, X.~Luo, T.~Hong.
\newblock Prototype energy models for data centers.
\newblock {\it Energy and Buildings\/} {\bf 231}, 110603 (2021).

\bibitem{dayarathna2015data}
M.~Dayarathna, Y.~Wen, R.~Fan.
\newblock Data center energy consumption modeling: A survey.
\newblock {\it IEEE Communications Surveys and Tutorials\/} {\bf 18}, 732 (2015).

\bibitem{crawford2024generative}
K.~Crawford.
\newblock {Generative AI’s Environmental Costs Are Soaring — and Mostly Secret}.
\newblock {\it Nature\/} {\bf 626}, 693 (2024).

\bibitem{erdenesanaa2023ai}
D.~Erdenesanaa.
\newblock {A.I. Could Soon Need as Much Electricity as an Entire Country}.
\newblock {\it The New York Times\/}  (2023). Sec. Climate.

\bibitem{patterson2021carbon}
{D. Patterson, J. Gonzalez, Q. Le, C. Liang, L-M. Munguia, D. Rothchild, D. So, M. Texier, J. Dean}.
\newblock {Carbon Emissions and Large Neural Network Training}.
\newblock {\it arXiv preprint arXiv.2104.10350\/}  (2021).

\bibitem{DataCenterMap24}
D.~C. Map, Data center map: Colocation, cloud, and connectivity, \url{https://www.datacentermap.com/} (2024). Accessed: 15-October-2024.

\bibitem{iea_datacentres}
{International Energy Agency}, {Data Centres and Data Transmission Networks} (2024). Accessed: 7-October-2024.

\bibitem{enerdata_electricityconsumption}
{Enerdata}, {Electricity Domestic Consumption Data} (2024). Accessed: 7-October-2024.

\bibitem{thenewstack_datacenters_energy}
{The New Stack}, {How Much Energy Is Really Being Consumed by Data Centers?} (2023). Accessed: 7-October-2024.

\bibitem{stokel-walker2024generative}
C.~Stokel-Walker.
\newblock The generative ai race has a dirty secret.
\newblock {\it Wired\/}  (2024). Accessed: 2-May-2024.

\bibitem{strubell2019energy}
E.~Strubell, A.~Ganesh, A.~McCallum.
\newblock {Energy and Policy Considerations for Deep Learning in NLP}.
\newblock {\it arXiv preprint arXiv.1906.02243\/}  (2019).

\bibitem{epa2024electricity}
{U.S. Environmental Protection Agency}, About the u.s. electricity system and its impact on the environment, \url{https://www.epa.gov/energy/about-us-electricity-system-and-its-impact-environment} (2024). Accessed: 7-October-2024.

\bibitem{siddik2021environmental}
M.~A.~B. Siddik, A.~Shehabi, L.~Marston.
\newblock The environmental footprint of data centers in the united states.
\newblock {\it Environmental Research Letters\/} {\bf 16}, 064017 (2021).

\bibitem{guidi2024environmental}
G.~Guidi, F.~Dominici, N.~Steinsultz, G.~Dance, {\it et~al.\/}.
\newblock The environmental burden of the united states' bitcoin mining boom.
\newblock {\it Research Square\/}  (2024). Preprint.

\bibitem{ibm_hyperscale_datacenter}
{IBM}, {Hyperscale Data Centers} (2024). Accessed: 7-October-2024.

\bibitem{iea_datacenters_networks}
{International Energy Agency}, {Data Centres and Data Transmission Networks} (2024). Accessed: 7-October-2024.

\bibitem{eia_energyconsumption}
{U.S. Energy Information Administration}, {How much electricity is consumed in the U.S. each year?} (2024). Accessed: 7-October-2024.

\bibitem{EIA_fuel_use}
{U.S. Energy Information Administration}, State energy data system (seds): Fuel use by state, \url{https://www.eia.gov/state/seds/sep_fuel/html/pdf/fuel_use_es.pdf} (2023). Accessed: 15-October-2024.

\bibitem{enerdata_domestic_consumption}
{Enerdata}, {Electricity Domestic Consumption Data} (2024). Accessed: 7-October-2024.

\bibitem{masanet2020recalibrating}
E.~Masanet, A.~Shehabi, N.~Lei, S.~Smith, J.~Koomey.
\newblock Recalibrating global data center energy-use estimates.
\newblock {\it Science\/} {\bf 367}, 984 (2020).

\bibitem{epa_egrid_faq}
{U.S. Environmental Protection Agency}, {Frequent Questions about eGRID} (2024). Accessed: 7-October-2024.

\bibitem{brander_most_2022}
M.~Brander.
\newblock The most important {GHG} accounting concept you may not have heard of: the attributional-consequential distinction.
\newblock {\it Carbon Management\/} {\bf 13}, 337 (2022). Publisher: Taylor \& Francis \_eprint: https://doi.org/10.1080/17583004.2022.2088402.

\bibitem{EIA_carbon_emissions}
{U.S. Energy Information Administration}, Carbon dioxide emissions from energy consumption, \url{https://www.eia.gov/environment/emissions/carbon/} (2023). Accessed: 15-October-2024.

\bibitem{EPA2023}
{United States Environmental Protection Agency}, Greenhouse gas inventory data explorer, \url{https://cfpub.epa.gov/ghgdata/inventoryexplorer/} (2023). Accessed: 17-October-2024.

\bibitem{ourworldindata_carbon_intensity}
{Our World in Data}, {Carbon Intensity of Electricity} (2024). Accessed: 7-October-2024.

\bibitem{henneman2023mortality}
{L. Henneman, C. Choirat, I. Dedoussi, F. Dominici, J. Roberts, C. Zigler}.
\newblock Mortality risk from united states coal electricity generation.
\newblock {\it Science\/} {\bf 382}, 941 (2023).

\bibitem{california_energy_commission}
C.~E. Commission, 2023 total system electric generation, \url{https://www.energy.ca.gov/data-reports/energy-almanac/california-electricity-data/2023-total-system-electric-generation}. Accessed: 28-October-2024.

\bibitem{google_sustainability2024}
2024 {E}nvironmental {R}eport - {G}oogle {S}ustainability, \url{https://sustainability.google/reports/google-2024-environmental-report/}. Accessed 29-October-2024.

\bibitem{microsoft_sustainability2024}
Microsoft, 2024 {E}nvironmental {S}ustainability {R}eport, \url{https://query.prod.cms.rt.microsoft.com/cms/api/am/binary/RW1lMjE}. Accessed 29-October-2024.

\bibitem{datacenter_knowledge_tougher_reporting}
D.~C. Knowledge, {T}ougher {R}eporting {M}andates {A}head for {D}ata {C}enters --- datacenterknowledge.com, \url{https://www.datacenterknowledge.com/regulations/tougher-reporting-mandates-ahead-for-data-centers}. Accessed 29-October-2024.

\bibitem{uptimeinstitute_mandates_crypto}
J.~Dietrich, {U}{S} mandates crypto energy reporting: will data centers be next? - {U}ptime {I}nstitute {B}log, \url{https://journal.uptimeinstitute.com/us-mandates-crypto-energy-reporting-will-data-centers-be-next/}. Accessed 29-October-2024.

\bibitem{bnef_clean_energy}
{C}orporate {C}lean {P}ower {B}uying {G}rew 12

\bibitem{xing2023carbonrespondercoordinatingdemand}
{J. Xing, B. Acun, A. Sundarrajan, D. Brooks, M. Chakkaravarthy, N. Avila, C-J. Wu, B. C. Lee}.
\newblock Carbon responder: Coordinating demand response for the datacenter fleet.
\newblock {\it arxiv preprint arxiv.2311.08589\/}  (2023).

\bibitem{jagannadharao2023timeshifting}
A.~Jagannadharao, N.~Beckage, D.~Nafus, S.~Chamberlin.
\newblock Timeshifting strategies for carbon-efficient long-running large language model training.
\newblock {\it Innovations in Systems and Software Engineering\/} pp. 1--15 (2023).

\bibitem{dodge2022measuringcarbonintensityai}
{J. Dodge, T. Prewitt, R. Tachet Des Combes, E. Odmark, R. Schwartz, E. Strubell, A. S. Luccioni, N. A. Smith, N. DeCario, W. Buchanan}.
\newblock Measuring the carbon intensity of ai in cloud instances.
\newblock {\it arXiv preprint arxiv.2206.05229\/}  (2022).

\bibitem{Gagnon_and_Cole2022}
P.~Gagnon, W.~Cole.
\newblock Planning for the evolution of the electric grid with a long-run marginal emission rate.
\newblock {\it iScience\/} {\bf 25}, 103915 (2022).

\bibitem{istrate2024environmental}
R.~Istrate, {\it et~al.\/}.
\newblock The environmental sustainability of digital content consumption.
\newblock {\it Nature Communications\/} {\bf 15}, 3724 (2024).

\bibitem{dante2024}
D.~Niewenhuis, S.~Talluri, A.~Iosup, T.~De~Matteis, {\it Companion of the 15th ACM/SPEC International Conference on Performance Engineering\/}, ICPE '24 Companion (Association for Computing Machinery, New York, NY, USA, 2024), p. 189–195.

\bibitem{sarkar2024carbon}
S.~Sarkar, {\it et~al.\/}, {\it Proceedings of the AAAI Conference on Artificial Intelligence\/} (2024), vol.~38, pp. 22322--22330.

\bibitem{shehabi2016united}
A.~Shehabi, {\it et~al.\/}.
\newblock United states data center energy usage report  (2016).

\bibitem{luers2024will}
A.~Luers, {\it et~al.\/}.
\newblock Will ai accelerate or delay the race to net-zero emissions?
\newblock {\it Nature\/} {\bf 628}, 718 (2024).

\bibitem{masanet2020much}
E.~Masanet, N.~Lei.
\newblock How much energy do data centers really use.
\newblock {\it Aspen Global Change Institute\/}  (2020).

\bibitem{cc_techgroup_2024}
{CC Tech Group}, How much power does a data center use per square foot? (2024). Accessed: 10-October-2024.

\bibitem{epa_egrid_2024}
{U.S. Environmental Protection Agency}, egrid - emissions and generation resource integrated database (2024). Accessed: 10-October-2024.

\end{thebibliography}

\noindent \textbf{Acknowledgments:} We benefited from helpful comments and suggestions from Michelle Audirac, Nat Steinsultz, and Henry Richardson. We wish to thank them for their support and insights.\\

\noindent \textbf{Funding:} National Institute of Environmental Health Sciences-Harvard Center Pilot Project P30ES000002 (FJBS). R01AG066793, R01ES034373. \vspace{1cm}\\
\noindent \textbf{Author contributions}
Conceptualization: FJBS, FD, GG, SD; 
Methodology: FJBS, FD, GG, SD; 
Data Collection and Data Wrangling: GG, KB, JG;
Investigation: FJBS, FD, GG, JG; 
Visualization: FJBS, FD, GG, KB; 
Funding acquisition: FJBS, FD; 
Project administration: FJBS, FD; 
Supervision: FJBS, FD;
Writing – original draft: FJBS, FD, GG;
Writing – review \& editing:  all authors.
\vspace{1cm}\\
\textbf{Competing interests:} The authors declare that they have no competing interests.
\vspace{1cm}\\
\textbf{Data and materials availability:} The dataset we constructed for our analysis will be made available for collaboration with the authors of this work. Requests can be addressed to the last author. Our code is hosted as an open-source project at: \\ \texttt{https://github.com/NSAPH-Projects/}.

\vspace{1cm}
\noindent \textbf{Supplementary Materials}\\
Materials and Methods\\
Robustness checks\\
Figs. S.1.1 to S.1.7\\
References (38-45)\\
\pagebreak

\appendix

\counterwithin{figure}{section}
\counterwithin{table}{section}
 \pagenumbering{arabic}
    \setcounter{page}{1}

\makeatletter
\def\@seccntformat#1{\@ifundefined{#1@cntformat}%
   {\csname the#1\endcsname\space}
   {\csname #1@cntformat\endcsname}}
\newcommand\section@cntformat{\thesection \space} 
\makeatother
\renewcommand{\thesection}{S.\arabic{section}}
\counterwithin{equation}{section}
\counterwithin{figure}{section}
\counterwithin{table}{section}
    
\begin{center}
    \Large \textbf{Supplementary Materials to: \\
    ``Environmental Burden of United States Data Centers in the Artificial Intelligence Era''}

    \vspace{0.25cm}
    \large Gianluca Guidi, Francesca Dominici, Jonathan Gilmour, Kevin Butler,\\ Eric Bell, Scott Delaney, Falco J. Bargagli-Stoffi
\end{center}

\doublespacing
\thispagestyle{empty}
\tableofcontents
\clearpage

\newpage

\pagenumbering{arabic}
\setcounter{page}{1}

\section{Materials and Methods}

To construct a comprehensive data platform for United States' (US) data centers, we implemented a multi-phase data pipeline involving data collection, validation, enrichment, prediction, and analysis. 

First, we acquired data on data centers from \textit{Baxtel.com}, a leading information resource for data centers in the US, to compile a foundational dataset. This dataset comprised 2,144 data centers across the US, and included provider name, exact location, and data center type for all data centers. Data center buildings' square footage and power capacities were available for only some of the data centers.  To enhance and validate this initial dataset, we integrated data obtained via web scraping from various online resources, which provided additional details and ensured accuracy, and in some cases filled in missing values for data center buildings' square footage. Further data validation and enrichment were achieved using satellite imagery from \textit{OpenStreetMap} (OSM), which allowed us to obtain precise measurements of data center buildings' square footage based on the geographical locations of the data centers.  Details on these first steps are provided in Section S.1.1.

Second, we estimated the power capacities for each data center in our dataset. Our initial dataset comprised 2,144 data centers across the United States, 2,132 of which were in the contiguous United States. For this study, we focused on data centers in the continguous US. For these data centers, we had data on power capacities for 1,795 data centers and had to estimate power capacities for the 337 missing data centers. We addressed the missing power capacity information using a Gradient Boosting Regression Tree (GBRT) model. This model utilizes features such as climate type, balancing authority, data center type (e.g., hyperscale or not), and building square footage to predict the power capacities accurately. Details on this second step are provided in Section S.1.2.

Third, we assigned each data center to their balancing authority region and its energy-supplying power plants. Post-prediction, each data center was assigned to its respective balancing authority region by matching their locations with the corresponding balancing authorities responsible for managing electricity supply and demand. Using data from the U.S. Environmental Protection Agency (EPA), we linked each data center to its supplying power plants. Details on this third step are provided in Section S.1.3.

Fourth, we calculated the energy demand of each data center and traced back their attributable carbon emissions based on the power plants' emission per MWh of load data, providing a comprehensive view of the environmental impact associated with each data center's energy consumption. Details on this fourth step are provided in Section S.1.4.
 
This multi-phase data pipeline facilitated the creation of a robust and comprehensive data platform for analyzing data centers in the United States, their energy consumption, attributable carbon emissions, and environmental impact. By integrating diverse data sources and employing advanced machine learning techniques, we were able to fill gaps in our dataset and conduct a detailed analysis of US data center power capacities and their environmental impact, while providing a valuable tool for stakeholders aiming to optimize data center operations and minimize their carbon footprint. 

\subsection{Data Collection}

\subsubsection{Baxtel Data}

The first part of the pipeline involved compiling a relevant dataset of data centers in the US. This dataset was created using multiple data sources and methodologies. Initially, we sought a reliable and comprehensive list of data centers in the US. Due to the absence of an open-source dataset providing this information, we acquired data from Baxtel, a leading information resource for data centers. Baxtel offers comprehensive insights into data centers and provides tools to support informed data center deployments. Baxtel supplied a dataset containing information on 2,144 data centers across the US. This dataset included the names of data center providers, addresses, latitudes, longitudes, and data center types for all 2,144 data centers. Power capacity data was available for 1,795 of these data centers, while square footage information was provided for 941 centers. For the analysis, we focused on the 2,132 data centers in the contiguous US.

All Baxtel's information is independently verified by Baxtel's expert researchers prior to inclusion in its database. The data is gathered through a diverse range of sources, including news outlets, media publications, RSS feeds, satellite imagery, direct communication with providers, and facility audits from some of the largest global interconnection providers.

\subsubsection{Web-Scraping}

The second step involved validating and enriching the initial dataset through web scraping from various data centers information providers websites. This effort resulted in a list of 1,182 data centers, including their provider names, complete addresses, countries, exact latitudes and longitudes, and square footage. The primary website scraped was \textit{datacenters.com}. 

The Baxtel dataset was then merged with our scraped dataset using latitude and longitude to ensure that no duplicates were included. This process allowed us to verify square footage information for data centers where we had data from both sources and to supplement the square footage details for Baxtel data centers that were missing this information.

\subsubsection{Open Street Map}

Finally, to further validate the obtained and collected data centers' facilities square footages, and enrich the dataset where information was still missing, we used Open Street Map. The methodology for determining the sizes of data center buildings involved several steps. We began by extracting all building data from the publicly available version of OpenStreetMap, which includes over 64 million buildings in the US. Next, we spatially intersected the geocoded addresses of the data centers in our dataset with these OpenStreetMap buildings, selecting those where the rooftop address fell within the building’s footprint. This process allowed us to determine the square footage for 1,587 out of the 2,132 data centers.

In total, we were able to obtain square footage information for 1,929 of the 2,132 data centers. When we had multiple estimates for the same data center, we prioritized Baxtel's data over the other estimates, as it was deemed the most accurate due to its more sophisticated and up-to-date retrieval methods.

Knowing the exact size of the facilities is crucial to calculating the power capacity estimate, as this information is not directly available.

\begin{figure}
\begin{center}
\includegraphics[width=1\textwidth]{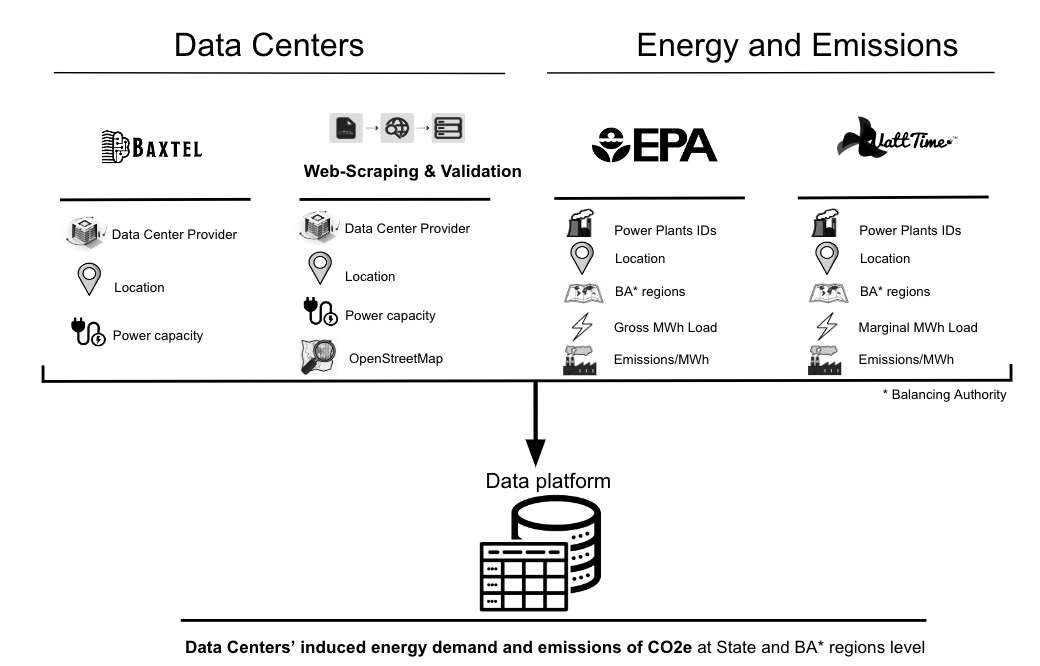}
\end{center}
\caption{\textbf{Data pipeline.} Data sources and types used to generate our data set.}
\label{fig:pipeline}
\end{figure}

\begin{figure}
\begin{center}
\includegraphics[width=1\textwidth]{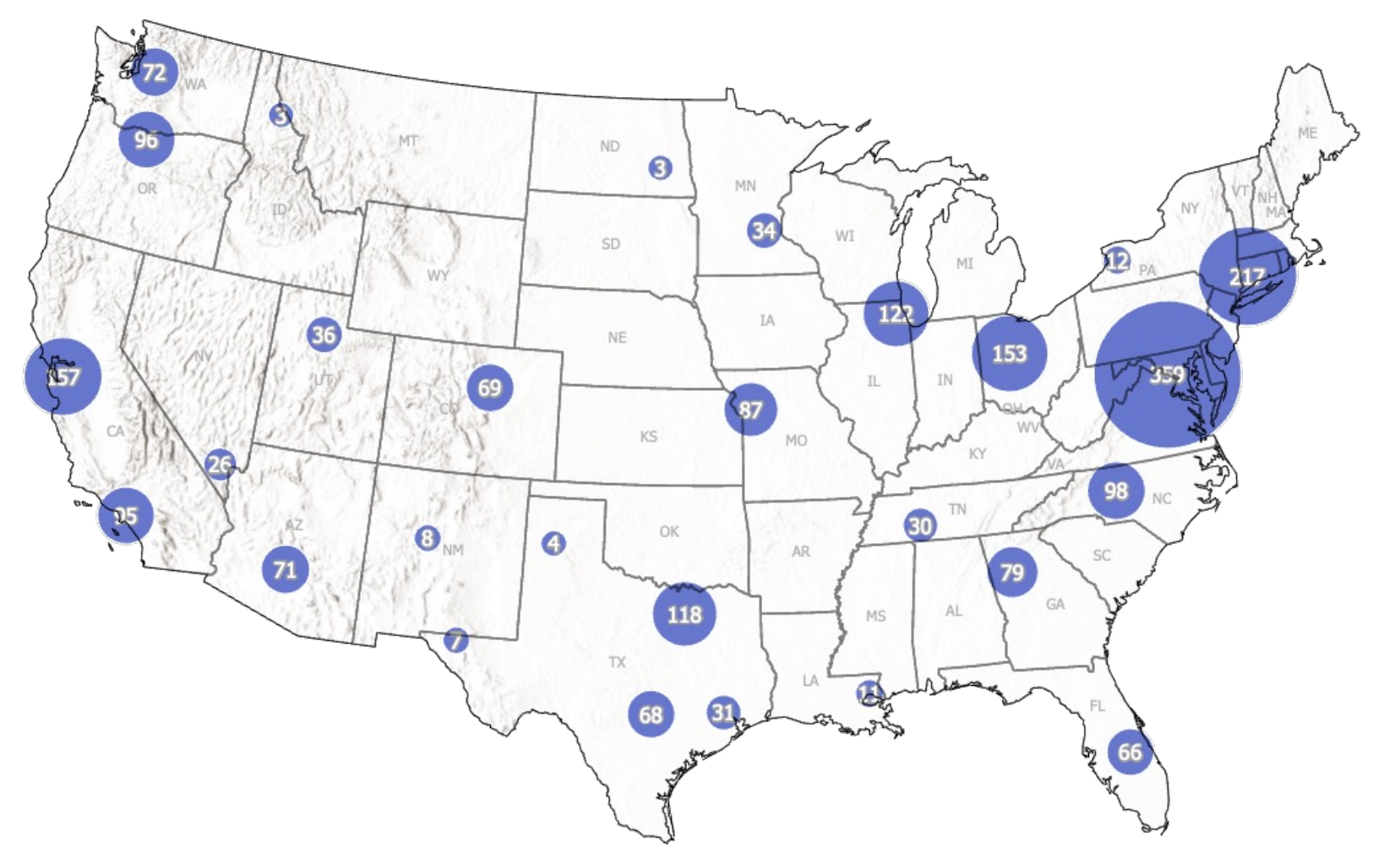}
\end{center}
\caption{\textbf{Geographic location of data centers in the contiguous US included in our analysis.} The figure depicts the 2,132 data centers in operation for the study period of September 2023 to August 2024. Circle size is proportional to the number of data centers in the geographic area.}
\label{fig:location}
\end{figure}

\subsection{Estimating Power Capacities}

In cases where power capacity data was missing, we estimated it using a machine learning approach. Specifically, we employed a GBRT model, leveraging a variety of features related to data center characteristics and their environments.

Our preliminary analysis, in line with existing literature, confirmed that the power capacity of a data center is highly correlated with its square footage \cite{istrate2024environmental, dante2024,sarkar2024carbon,masanet2020recalibrating,shehabi2016united,luers2024will,siddik2021environmental,masanet2020much}. By excluding outliers (defined as data points with a Z-score greater than 2.0), we found an average power density of approximately 91.75 watts per square foot. This figure is relatively conservative compared to the 150 watts per square foot average reported in the literature, and indicates that our results provide a lower bound to the actual energy consumption and emissions levels \cite{cc_techgroup_2024}. 

The features used in our model included:
\begin{itemize}

\item Climate Type: The climate category of the region where the data center is located, as different climates can affect cooling requirements and, consequently, power consumption. This information was retrieved from Nasa Earth data, at the link \url{https://webmap.ornl.gov/ogcdown/dataset.jsp?dg_id=10012_1$}.

\item Balancing Authority: The entity responsible for ensuring a balance between electricity supply and demand in the region, which can influence the operational characteristics of data centers.

\item Data Center Type: A categorical variable indicating whether the data center is a hyperscale facility or not. Hyperscale data centers often have different design and operational standards compared to smaller facilities. For the model, all non-hyperscale data centers were grouped together. 

\item Square Footage: The total building square footage of the data center, a critical factor in determining power requirements.

\end{itemize}
To predict the missing power capacities, we split the data that included all known power capacities into a training set and a testing set. 

We used the following preprocessing steps:
Scaling and Encoding: Numerical features were standardized to have a mean of zero and a standard deviation of one. Categorical features were converted into one-hot encoded vectors to allow the model to interpret them correctly.

The Gradient Boosting Regressor was then fitted to the training data. Gain-based feature importance scores were extracted to understand the contribution of each feature to the model's predictions. Amongst all the features, data center square footage was the most important (0.76) factor in predicting the data center's power capacity. The data center type was also relevant in the prediction, specifically when the data center was a hyperscale facility (0.11).
The performance of our model was evaluated using the R-squared metric and the mean test error. The model achieved an R-squared value of 0.77, indicating a high potential of our model to explain the power capacity given the independent variables. Moreover, the mean test error, computed as the difference between the actual values and the predicted values, was positive, suggesting that our model tended to underestimate the power capacities on average, yielding thus conservative estimates.

Our GBRT model proved to be a robust approach for estimating the power capacity of data centers based on their physical and operational characteristics. 
This model was then utilized to predict power capacities for data centers where this information was missing.

\begin{figure}
\begin{center}
\includegraphics[width=0.6\textwidth]{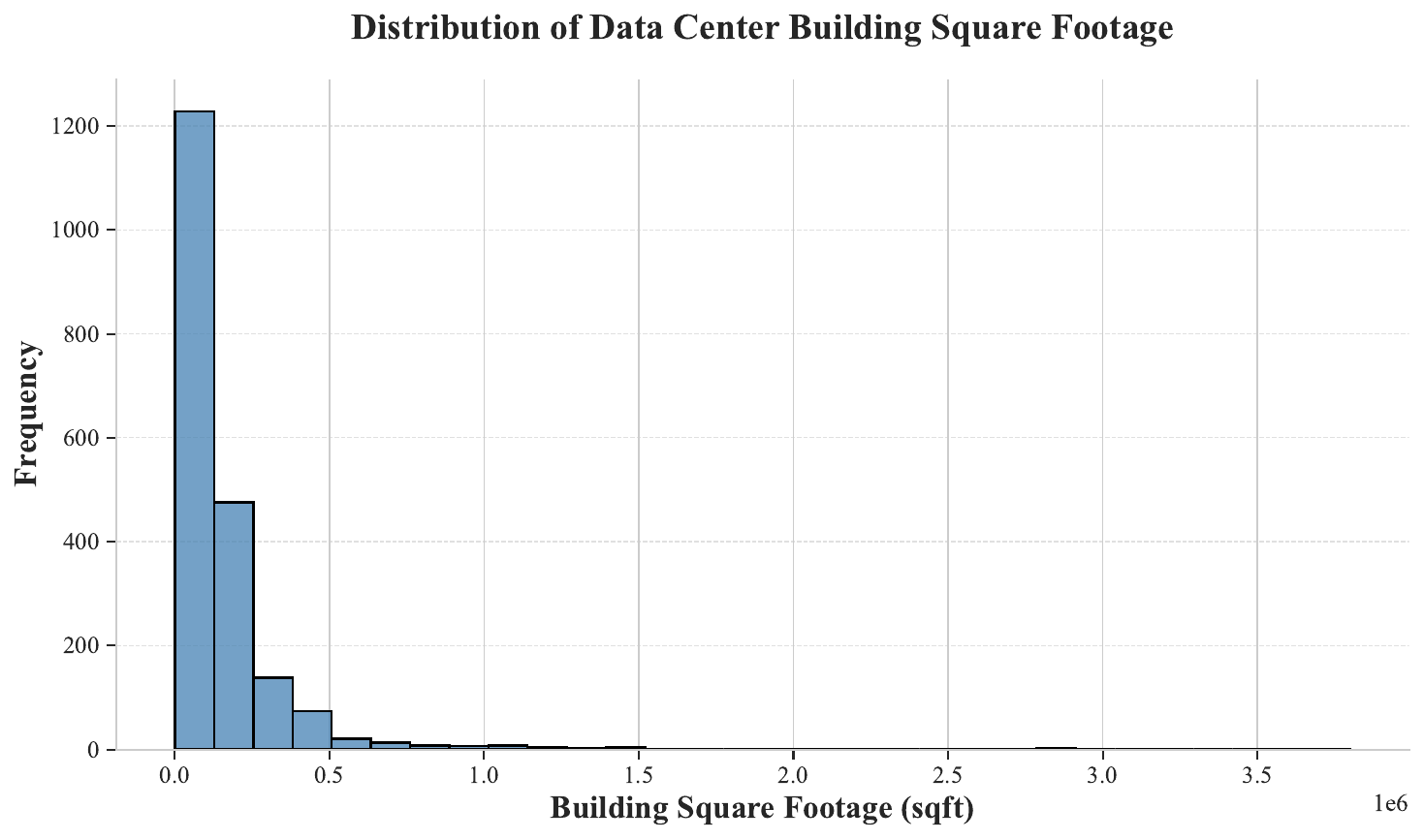}
\end{center}
\caption{\textbf{Data Center Square Footage Distributions.}}
\label{fig:}
\end{figure}

\begin{figure}
\begin{center}
\includegraphics[width=0.6\textwidth]{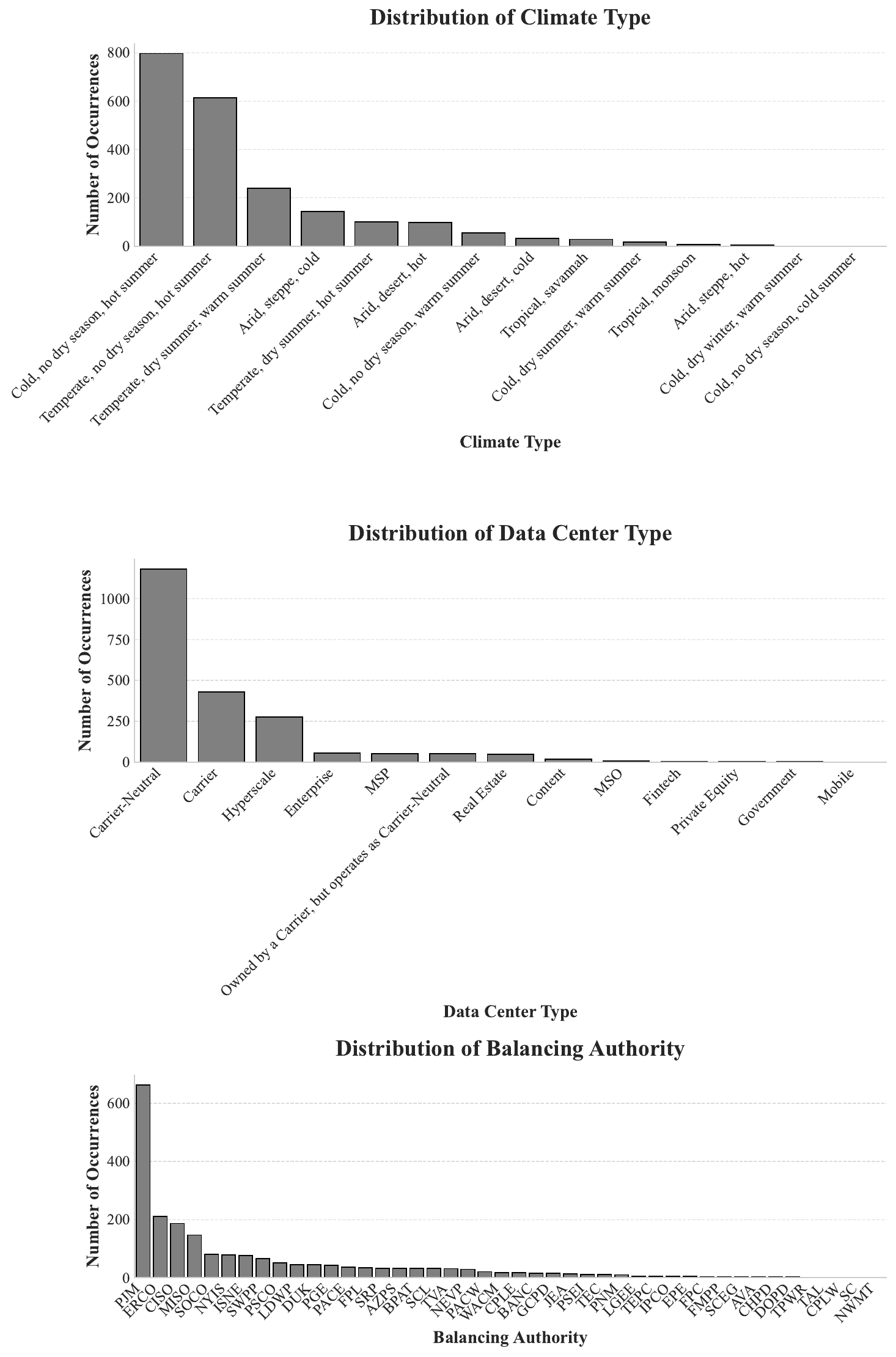}
\end{center}
\caption{\textbf{Data Center Features Categorical Distributions.}}
\label{fig:}
\end{figure}

\begin{figure}
\begin{center}
\includegraphics[width=0.7\textwidth]{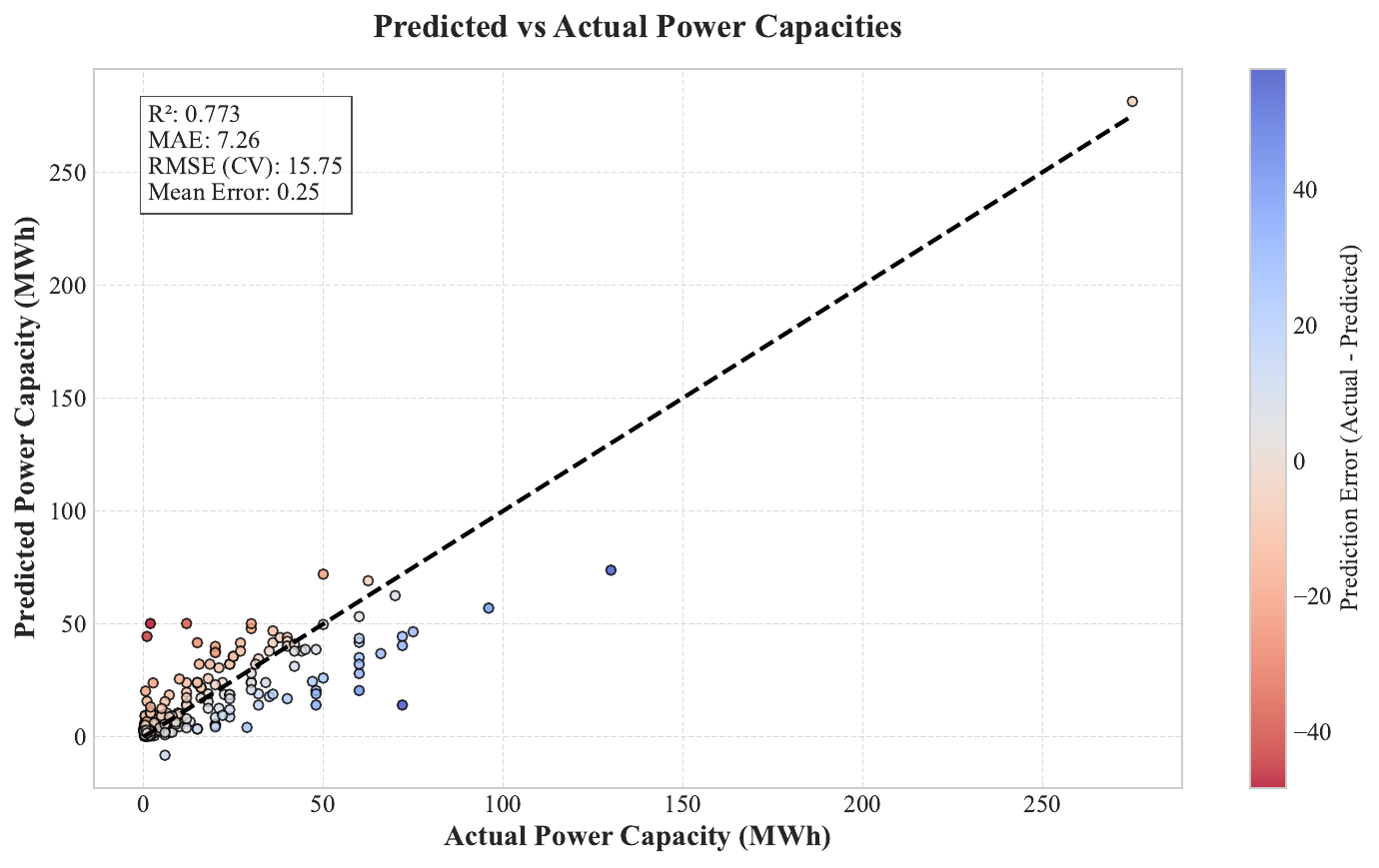}
\end{center}
\caption{\textbf{Data Center Power Capacity Estimation Model Performance.}}
\label{fig:}
\end{figure}

\begin{figure}
\begin{center}
\includegraphics[width=0.6\textwidth]{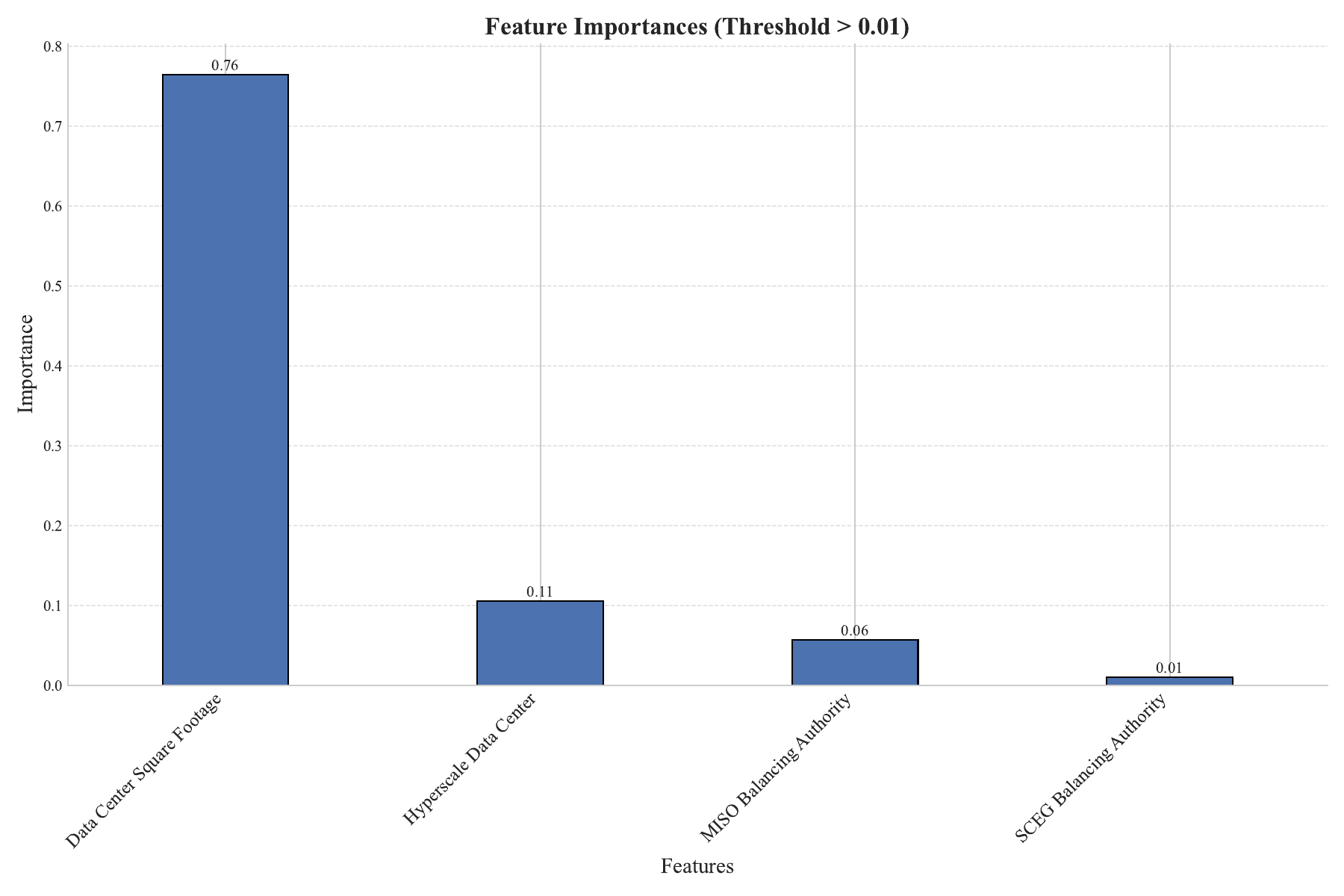}
\end{center}
\caption{\textbf{Data Center Features Importance to Power Capacity Model.}}
\label{fig:}
\end{figure}

\subsection{Balancing Authorities and Power Plants} 

To connect each data center to its supplying power plants, we utilized the EIA's balancing authority (BA) regions. These geographic polygons, provided by WattTime—a nonprofit focused on tracking greenhouse gas emissions—offer detailed coverage of power generation and supply. WattTime gathers data from multiple sources, including the U.S. Environmental Protection Agency (EPA), grid emissions, satellite imagery, and machine learning models.
Using these BA regions, we matched both data centers and power plants to their respective BA regions based on location. From there, we accessed power plants' emissions coefficients per megawatt-hour (MWh) from the EPA's Emissions and Generation Resource Integrated Database (eGRID).
Finally, we calculated the energy contribution of each power plant supplying a data center by developing a weighted model based on energy generation. Further details on this model are provided in the next section.

\begin{figure}
\begin{center}
\includegraphics[width=1\textwidth]{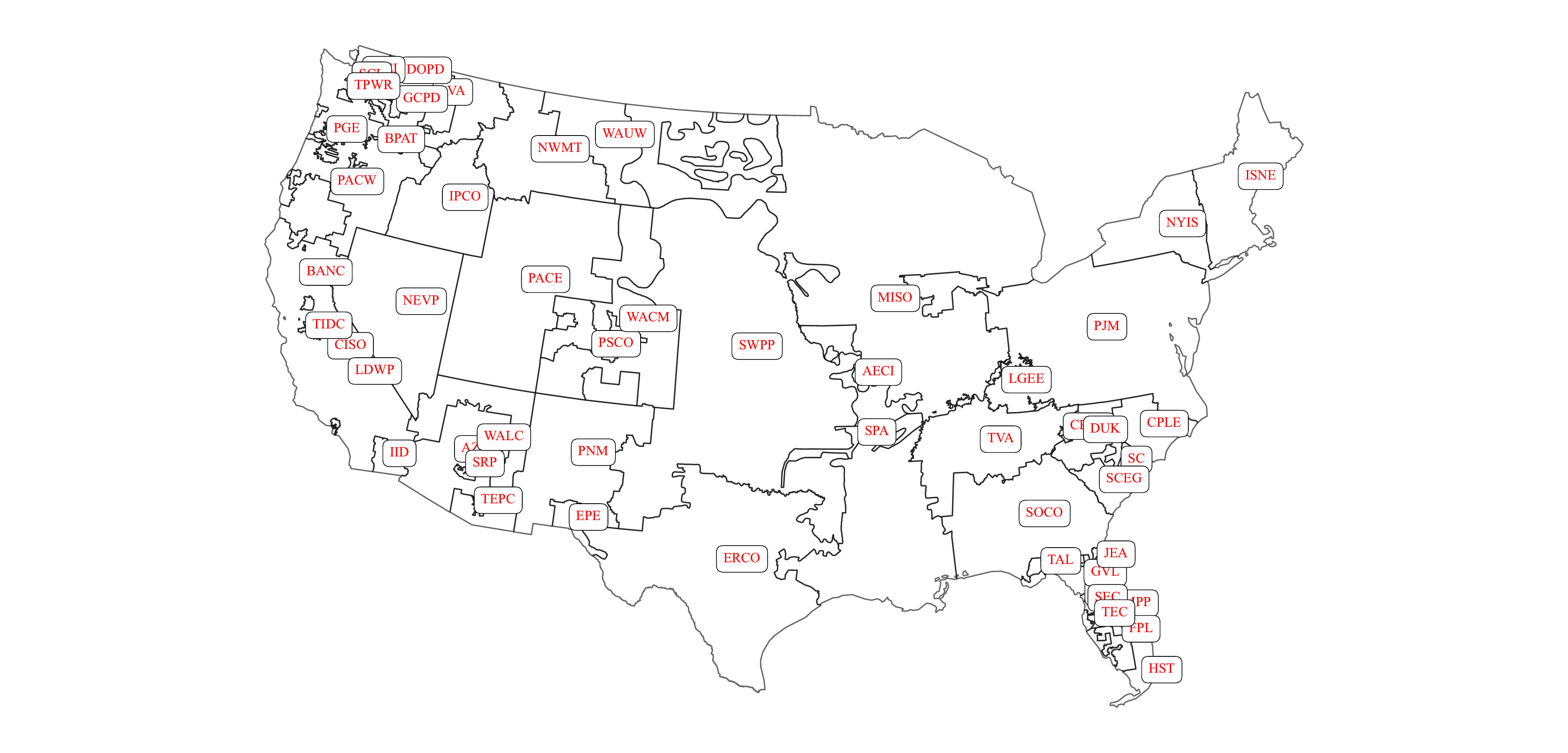}
\end{center}
\caption{\textbf{Balancing Authorities and Regions in the US.} }
\label{fig:BA}
\end{figure}
\subsection{Computing Data Center Energy Consumption, Attributable Carbon Emissions, and Carbon Intesity}

The last part of the pipeline consisted of computing the annual energy load of each data center, distributing this energy load amongst its supplying power plants, computing the CO$_2$e emissions related to those energy load supplies, and calculating the carbon intensity. 

\subsubsection{Data Centers' Energy Load}

To compute each data center energy load, we multiplied its power capacity by 8,760 hours/year, representing the total number of hours in a year. 
Finally, we assumed that the uptime is set for all data centers to 75\%, meaning the data centers operate continuously throughout the entire year without any downtime. Therefore, the computation reads as:

\begin{equation}
\text{Data Center Energy Load (i, y)} = \text{Power Capacity}_{i} \cdot 8,760 \text{ hours/year} \cdot \textit{uptime}
\end{equation}

where:
\begin{itemize}
    \item $\text{Data Center Energy Load (i, y})$ is the total energy load for data center $i$ in year $y$, measured in kilowatt-hours (kWh).
    \item $\text{Power Capacity}_{i}$ is the power capacity of data center $i$, measured in kilowatts (kW).
    \item \textit{uptime} refers to a coefficient that ranges from 0 to 1, representing the proportion of time a data center is assumed to be operational at full capacity. A value of 1 indicates the data center was fully operational 100\% of the time, while a value closer to 0 suggests it was active for only a small fraction of the time. This coefficient quantifies the actual usage or workload of the data center over the year.
\end{itemize}

By considering a 75\% uptime, this calculation provides an estimate assuming the data center operates at full capacity for 75\% of the year.



\subsubsection{Energy Generation Weighted Model}

In order to allocate each power plant within a BA region its respective share of the energy load stemming from the energy demand of the data centers in that region, we developed an energy generation weighted model (EGW), that computes the energy load for each power plant and ultimately estimates the corresponding emissions. 

The underlying assumption of the EGW model is that within a BA, each power plant supplies energy proportional to the total energy it generated over the year. The coefficient of energy production contribution was calculated as follows:

\begin{equation}
COEFF_{EGW}(y, j, B) = \frac{\text{Power Plant Annual Net Energy Generation}(y, j, B)}{\sum_{j=1}^{N_B} \text{Power Plant Annual Net Energy Generation}(y, j, B)}
\end{equation}





where:
\begin{itemize}
    \item $y$ represents the year of analysis.
    \item $j$ denotes the index of the power plant within the balancing authority.
    \item $B$ stands for the balancing authority.
    \item $N_B$ is the total number of power plants within the balancing authority $B$.
    \item $\text{Power Plant Annual Net Energy Generation}(y, j, B)$ is the total energy generated by power plant $j$ in year $y$ within balancing authority $B$.
\end{itemize}

This coefficient represents the fraction of the total annual energy generation attributed to each power plant within a BA. By definition, the sum of these coefficients for all power plants in a BA equals one:

\[
\sum_{j=1}^{N_B} COEFF_{EGW}(y, j, B) = 1
\]

Data for these computations were retrieved from the EPA 2022 eGrid \cite{epa_egrid_2024}, for all 3,318 power plants above 25 MW capacity that reported their energy annual net generation of energy and emissions of CO$_{2}$e per MWh of load. 

Next, we calculated the attributional load assigned to each power plant using the EGW approach. The power plant's load was determined by multiplying the total additional energy demand from data centers by the plant's EGW coefficient:

\begin{equation}
\text{Power Plant Load}_{EGW}(i, y, j, B) = COEFF_{EGW}(y, j, B) \cdot \text{Data Center Energy Load (i, y, B)}
\end{equation}

where:
\begin{itemize}
    \item $i$ denotes the index of the data center causing the additional load.
    \item $\text{Data Center Energy Load (i, y, B)}$ represents the additional energy demand from data center $i$ in year $y$ within balancing authority $B$.
\end{itemize}

To estimate the emissions attributed to the load of each power plant, we applied the average emission factors specific to each plant. These emission factors are typically expressed as emissions per unit of energy generated (e.g., pounds of CO$_2$ per MWh). The emissions for each power plant were then computed as:

\begin{equation}
\text{Power Plant Emissions}_{EGW}(i, y, j, B) = \text{Power Plant Load}_{EGW}(i, y, j, B) \cdot \text{Emission per MWh}(y, j, B)
\end{equation}

where:
\begin{itemize}
    \item $\text{Emission per MWh}(y, j, B)$ is the annual average CO$_2$ emissions factor for power plant $j$ in year $y$ within balancing authority $B$.
\end{itemize}

This approach allowed us to accurately attribute emissions to each power plant based on its proportional contribution to the additional energy demand from data centers. The calculated emissions take into account both the amount of energy each power plant generates and the specific emissions profile of each plant, providing a detailed and precise estimate of the attributional environmental impact of data center energy consumption.

\subsubsection{Average Carbon Intensity}
To quantify the efficiency of data centers' energy sourcing, we introduced a coefficient called carbon intensity of electricity generation attributed to data centers. Average carbon intensity was defined as the ratio of total carbon dioxide equivalents (CO$_{2}$e) emissions produced to the total energy generated by the power plant, expressed in grams of CO$_2$e per kilowatt-hour (gCO$_2$e/kWh).
The carbon intensity of a power plant was calculated using the following formula:

\begin{equation}
\text{Carbon Intensity}(i, y, j, B) = \frac{\text{Power Plant Emissions}_{EGW}(i, y, j, B) }{\text{Power Plant Load}_{EGW}(i, y, j, B)}
\end{equation}

where:
\begin{itemize}
    \item $\text{Carbon Intensity}(i, y, j, B)$ is the carbon intensity of power plant $j$ for data center $i$ in year $y$ within balancing authority $B$.
    \item $\text{Power Plant Emissions}_{EGW}(i, y, j, B)$ represents the total CO$_2$e emissions (in grams) from power plant $j$ attributed to the energy demand of data center $i$ in year $y$ within balancing authority $B$.
    \item $\text{Power Plant Load}_{EGW}(i, y, j, B)$ is the energy generated by power plant $j$ (in kWh) to meet the additional demand from data center $i$ in year $y$ within balancing authority $B$.
\end{itemize}

The carbon intesity was computed at various geographical levels, from balancing authorities to the entire country. 
The formula for carbon intensity at the balancing authority level is:

\begin{equation}
\text{Carbon Intensity}(B, y) = \frac{\sum{j=1}^{N_B} \text{Power Plant Emissions}_{EGW}(y, j, B) }{\sum{j=1}^{N_B} \text{Power Plant Load}_{EGW}(y, j, B)}
\end{equation}

where:
\begin{itemize}
\item $\text{Carbon Intensity}(B, y)$ is the carbon intensity for balancing authority $B$ in year $y$, expressed in grams of CO$_2$e per kilowatt-hour (gCO$2$/kWh).
\item $\text{Power Plant Emissions}_{EGW}(y, j, B)$ is the total CO$_2$ emissions (in grams) from power plant $j$ within balancing authority $B$ in year $y$.
\item $\text{Power Plant Load}{EGW}(y, j, B)$ is the energy generated by power plant $j$ (in kWh) to meet the additional demand from data centers in balancing authority $B$ during year $y$.
\item $N_B$ is the total number of power plants within the balancing authority $B$.
\end{itemize}

\section{Consequential Analysis}

We also conducted a consequential analysis of our carbon emissions results using data from WattTime. In the consequential analysis, we estimated the total emissions using the marginal power plants' emissions resulting from the WattTime's methodology, extensively presented in \cite{guidi2024environmental}. 
\subsection{Marginal Emissions Rate Computation and Results}

\noindent \textbf{Results.} Following this consequential approach, the total CO2 emissions caused by the 2132 data centers in our sample, keeping the uptime at 0.75, corresponded to 101.21 MT CO$_{2}$.





\end{document}